\documentclass[%
 reprint,
 amsmath,amssymb,
 aps,
pra,
]{revtex4-2}

\usepackage{tabularx}
\usepackage{dcolumn}
\usepackage{bm}
\usepackage{amsmath} 
\usepackage{adjustbox}
\usepackage{xcolor}
\usepackage{subfigure}
\usepackage{graphicx}
\usepackage{hyperref}
\usepackage{placeins}  
\usepackage{physics}
\usepackage{mathtools}
\begin{document}

\preprint{APS/123-QED}

\title{Quantum neural networks facilitating quantum state classification}
\author{Diksha Sharma$^{1}$}
\email{sharma.49@iitj.ac.in}
\author{Vivek Balasaheb Sabale$^{1}$}
\email{sabale.1@iitj.ac.in}
\author{Thirumalai M$^{1}$}
\email{m23iqt008@iitj.ac.in}
\author{Atul Kumar$^{1}$}
\email{corresponding author: atulk@iitj.ac.in}

\affiliation{$^{1}$ Indian Institute of Technology Jodhpur, 342030, India}

\begin{abstract}

The classification of quantum states into distinct classes poses a significant challenge. In this study, we address this problem using quantum neural networks in combination with a problem-inspired circuit and customised as well as predefined ans\"{a}tz. To facilitate the resource-efficient quantum state classification, we construct the dataset of quantum states using the proposed problem-inspired circuit. The problem-inspired circuit incorporates two-qubit parameterised unitary gates of varying entangling power, which is further integrated with the ans\"{a}tz, developing an entire quantum neural network. To demonstrate the capability of the selected ans\"{a}tz, we visualise the mitigated barren plateaus. The designed quantum neural network demonstrates the efficiency in binary and multi-class classification tasks. This work establishes a foundation for the classification of multi-qubit quantum states and offers the potential for generalisation to multi-qubit pure quantum states.
\end{abstract}

\maketitle

\section{Introduction}
Quantum machine learning \cite{biamonte2017quantum,schuld2015introduction,SHARMA2023128938,zhang2020recent} (QML) has seen substantial interest from researchers and industries across various domains such as finance, medicine, weather forecasting, drug discovery, image processing, cybersecurity, and many more \cite{armaos2020computational,rebentrost2024quantum,fedorov2021towards,cong2019quantum,guo2022quantum,sarkar2025quantum,zeema2025exploring,suhas2023quantum}. In general, QML focuses on enhancing classical machine learning \cite{jordan2015machine,bishop2006pattern} by leveraging quantum hardware, designed to exploit the quantum mechanical properties of quantum systems. In a nutshell, QML integrates quantum computing with classical machine learning to enhance the theoretical and mathematical insights into machine learning tasks for harnessing the unique strengths of both paradigms. This amalgamation of classical machine learning and quantum computing efficiently resulted in analyzing complex problems, such as quantum measurements, assessing, controlling, and simulating quantum systems, exploring boundaries between quantum phases, symmetry breaking, and identifying different types of phase transitions \cite{hu2017discovering,broecker2017machine,canabarro2019unveiling,kottmann2020unsupervised,flurin2020using,krenn2023artificial}. For quantum many-body systems, machine learning algorithms are utilized to predict the exact ground states of quantum many-body Hamiltonians \cite{huang2022provably}. Clearly, computing the exact ground state energy lies at the heart of condensed matter \cite{bedolla2020machine}, quantum simulation \cite{hu2020quantum, PhysRevA.106.022424}, and quantum chemistry \cite{dral2020quantum}. \par
From the perspectives of quantum computing- both foundational and applications- one of the key challenges is the classification of entangled systems, which aims to address the entanglement versus separability paradigm of quantum states \cite{PhysRevA.65.032314,Schmidt_number,horodecki1996separability, PhysRevLett.77.1413, PhysRevA.61.042314}. Classifying quantum states as entangled or separable is crucial, as entanglement and non-local correlations are fundamental to various quantum protocols \cite{bennett_brassard_1984,faujdar2021comparative} and real-world applications \cite{ecker2023advances,liao2017satellite}. The challenge becomes much more intricate as we move from two qubits to three qubits to multiqubit systems as the number of entangled classes keeps increasing with the increasing number of qubits \cite{dur2000classification,verstraete2002four}. \par

\begin{figure}[t!]
    \centering
    \includegraphics[width=\linewidth]{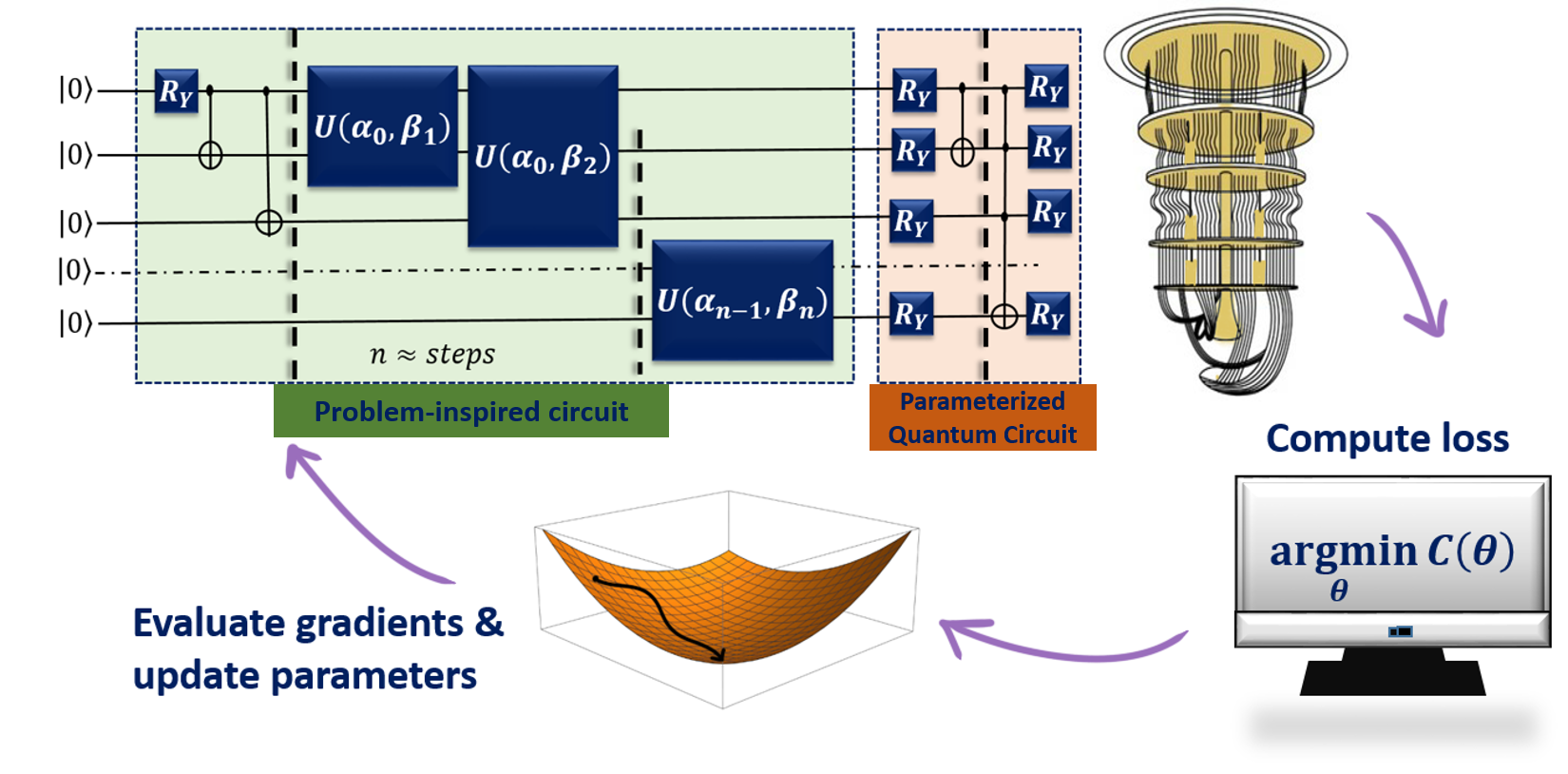}
    \caption{A general framework utilized for quantum neural networks for classifying the quantum state into entangled and separable classes.}
    \label{fig:flow_chart}
\end{figure}
Machine learning algorithms offer a promising approach to address this complex problem. Recent studies have applied
classical neural network algorithms to analyze entanglement versus separability paradigm in three-qubit quantum systems. For example, \cite{gulati2024ann} employed an artificial neural network (ANN) to classify three-qubit pure states as entangled or separable. For this, the authors used density matrix elements as features and reduced the required 128 features to 18 essential features. They further demonstrated the efficiency of the algorithm for classifying states as separable, bi-separable or entangled based on number of features and achieved a maximum accuracy of 85.4\% with 6 number of features. On similar lines, a deep convolution neural network and Siamese network are utilized to identify the bipartite entanglement in three-qubit states, where the dataset is generated using random density matrices \cite{pawlowski2024identification}. The maximum accuracy achieved using the algorithm was 98.31\%. Due to the complex nature of quantum states, additional features were required to incorporate the real and imaginary parts, advocating a need for such complex-to-real feature approximation in classical machine learning algorithms. Further, a study presented in \cite{asif2023entanglement}, formulates Bell-type inequalities using relative entropy of coherence and encodes it into an artificial neural network to classify pure quantum states.  In addition, some studies have utilised other machine learning algorithms, mainly support vector machines \cite{sharma2025harnessing, vintskevich2023classification} and bagging-based models \cite{lu2018separability}. Few studies approached the entanglement classification problem using quantum algorithms. \cite{qiu2019detecting} leveraged deep quantum neural network for classifying the quantum states into entangled or separable. However, the algorithms utilised multiple hidden layers for classification purposes. Another study \cite{sharma2025harnessing} employed a quantum support vector machine for classifying two-qubit mixed states. These studies have utilised multiple qubits and classified the states into two main classes \textit{i.e.} entangled or separable states. However, an increase in possible classes corresponding to the size of a system necessitates the use of multi-class classification techniques.\par
% This approximation may be overcome in QML algorithms, and the present study represents a significant step in this direction. 
% The trained model is further utilized to evaluate the positive-partial-transposition entangled states.
QML algorithms are definitely a significant step in this direction to address such issues, which use the strengths of both machine learning and quantum computing to efficiently resolve entanglement versus separability paradigm in multiqubit systems. In this article, we present the use of quantum algorithms, specifically quantum neural networks (QNN). To analyse the quantum state using QNN, the required training data is generated using a quantum circuit that constructs the quantum states with required classes. This circuit is utilized as a feature mapping circuit that constructs the quantum states and is fed to the QNN algorithm. A general framework of the QNN is shown in Fig. \ref{fig:flow_chart}. We further present the scalability of our model by demonstrating the classification of three-qubit quantum states in five possible classes, also providing the necessary tools to attempt a multi-class classification of multi-qubit quantum states. The model demonstrates the generation of quantum states utilising a minimum number of qubits and uses the generated states for classification purposes without explicitly storing the states. The use of such a circuit also reduces the required resources for running QNN and other QML algorithms.  \par

\section{Preliminaries}
In this section, we cover basic concepts such as Entanglement and the workings of QNN. Section \ref{ent} aims to provide the necessary understanding of quantum entanglement, and section \ref{NN} covers the workings of QNN. 
% We give a brief introduction of these concepts.
\subsection{Entanglement\label{ent}}
The composite quantum systems can exist in correlated quantum states and exhibit non-classical behaviour known as entanglement. Quantum entanglement is a resource for various quantum information protocols and quantum computing. Understanding, utilising, and quantifying the available resources is one of the main focuses of quantum information. Quantifying entanglement present in quantum states is a challenging task, in addition to classifying and identifying various entanglement classes of quantum states. The possible classes for a bipartite system (A $\&$ B) are entangled (AB) and separable (A-B); one can rely on Peres-Horodecki criteria \cite{HORODECKI19961}, Schmidt number \cite{Schmidt_number}, or concurrence \cite{PhysRevLett.77.1413} to determine the quantum states class.  In the case of a multi-partite quantum state, the possible entanglement classes are very high and complex to identify. The problem of assigning a class for a simple case of three-qubit quantum states is also difficult due to the increase in total classes (ABC, A-BC, B-AC, C-AB, A-B-C) \cite{PhysRevA.62.062314, PhysRevA.61.042314}. The entropy of all possible single qubit density matrices resulting from a partial trace operation performed on a density matrix corresponding to many qubit systems is used to assign classes. The process of taking partial trace is difficult and computationally costly. We use QNN to address this difficulty and reduce computational cost. We also provide the necessary tools for scaling this process in many-qubit pure quantum states classification.

\begin{figure*}[t!]
    \centering
    \includegraphics[width=\linewidth]{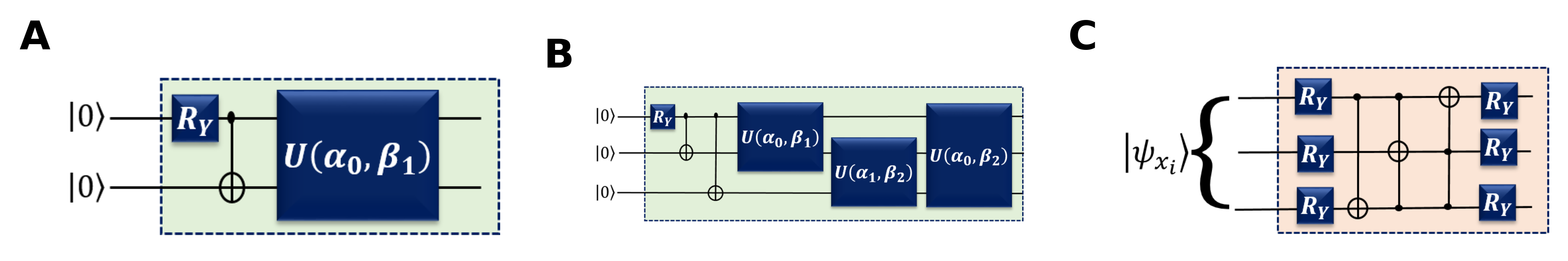}
    \caption{The problem-inspired circuits utilized for generating the desired quantum states A) correspond to the circuit for two-qubit quantum states generation, B) represent the circuit for three-qubit quantum states generation, and C) the proposed quantum circuit utilized as PQC in three-qubit quantum state classification.}
    \label{fig:combine_circuits}
\end{figure*}
 
\subsection{Quantum Neural Network\label{NN}}
A QNN is a hybrid quantum-classical algorithm utilised to solve various complex problems in QML. For completeness, we first introduce the QNN architecture. Suppose we have access to a set of datasets with $n$-samples consisting of feature vectors ${x_{i}}$ and corresponding class/label $y_{i}$, represented as $D=\{x_{i},y_{i}\}$. The dataset is partitioned into training and testing states in order to make the machine learning model learn and accurately evaluate unseen datasets. A QNN consists of a feature mapping circuit and a parameterised quantum circuit, which are optimised using a classical optimiser to make the model learn and assign the labels $y$ to appropriate $x$. Although all of these components of QNN play an important role, the feature mapping circuit utilised to properly encode the provided feature vectors $x_{i}$ into quantum states $\ket{\psi_{x_i}}=U_{x_{i}}\ket{0}$, where $U(x_{i})$ is a unitary gate to encode each dimension of the feature vector $U(x_{i})= e^{-ixF}$. Here, $x$ is the classical value, and $F$ is the required Hermitian matrix. The unitary circuit generates an encoded input state. Further, the encoded input data states are fed to a PQC, also called Ans\"{a}tz, which is denoted as $U_{\theta}$, resulting in the transformation of the input state
\begin{equation}
   \ket{\psi_{x_{i}}(\theta)}= U_{\theta} \ket{\psi_{x_i}}.
\end{equation}
Overall, the state can be described as $\ket{\psi_{x_{i},\theta}} = U_{\theta}U_{x_{i}}\ket{0}$. In addition, a parameterised quantum circuit ($U_{\theta}$) can be applied sequentially in $L$ numbers, classically equivalent to several hidden layers, represented as
\begin{equation}
    U_{\theta} =U_{\theta_L} U_{\theta_{L-1}} \cdots U_{\theta_{1}}
\end{equation}
where $\theta = \{\theta_1, \theta_2, \dots, \theta_L\}$ is the set of parameters to be optimised using classical optimizers. 
Later on, an expectation value is computed for a measurable observable ($O$), which is further utilised to compute the cost function of the QNN.
\begin{equation}
    \ket{\psi_{x_{i},\theta}} = \bra{0}U^{\dagger}_{x_{i}}U^{\dagger}_{\theta_{i}}OU_{x_{i}}U_{\theta_{i}}\ket{0}
\end{equation}
where the measurable operator maps the state to a scaler number, which can be considered as the predicted class/label. For better classification accuracy, a minimum cost/loss function is computed, numerically represented as $C(\theta) = argmin_{\theta}(y_i - \bra{\psi_{x_i,\theta}}O\ket{\psi_{x_i,\theta}})$, where $y_i $  is the true classification. The observable $O= I - \ket{0}\bra{0}^{\otimes n}$, where $n$ represents the projection operator on $n$ qubit space and is used to compute the cost function. The exploration of cost function parameter space may land us on the flat region in the cost landscape, known as a barren plateau, where the cost function gradient is almost zero. The presence of a barren plateau implies a reduction in the algorithm's trainability. We explored the possibility of such regions for a given neural network. Further, we demonstrate the benefits of using the proposed customised an\"{s}atz.

\begin{figure*}[t!]
    \centering
    \includegraphics[width=\textwidth]{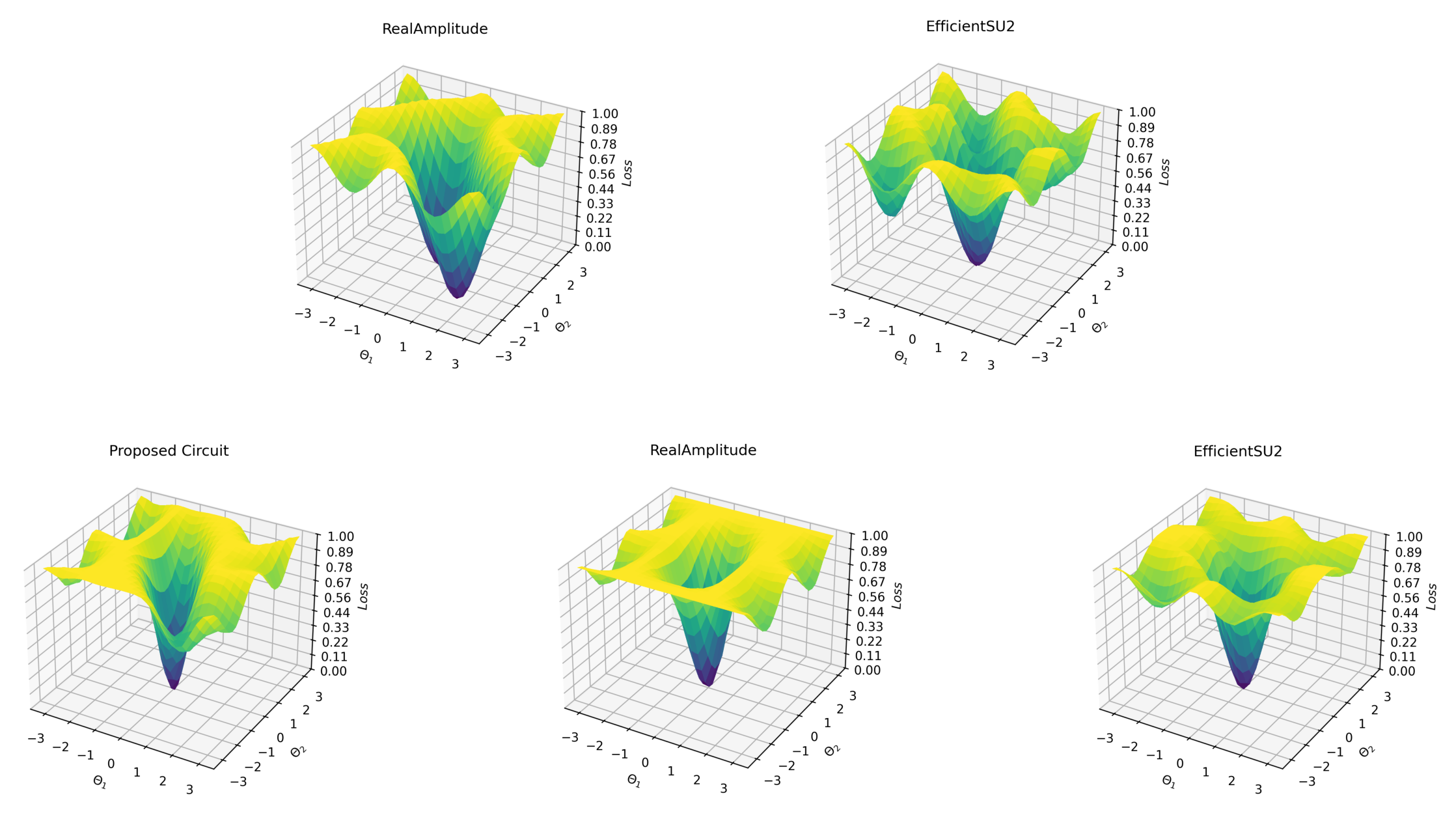}
    \caption{The barren plateau landscape for quantum state classificational models. The top row corresponds to two-qubit parameterised quantum circuits, and the bottom row corresponds to three-qubit parameterised quantum circuits.}
    \label{fig:barren_plateaus}
\end{figure*}

\section{Methodology}

The most crucial point in classification is to generate a well-balanced and diversified dataset. In general, the feature map circuit in a QNN accepts the data points from a stored dataset and constructs quantum states, which are transferred to a PQC. The expectation value associated with PQC is estimated using a quantum processor. These values further need to be optimised by using classical optimisers and re-optimising the parameters with the help of classical optimizers. On the basis of updated parameters, the loss is calculated to generalise the model on unseen datasets. To classify the quantum states into entangled or separable, researchers have utilised density matrix formalism to generate pure quantum states, which require many qubits for representation. On the other hand, utilising a circuit to generate the quantum states requires different circuits to generate both entangled and separable quantum states before employing the QML algorithms in entangled state classification. Therefore, we propose a model inspired by a QNN to classify the quantum states, represented in Fig. \ref{fig:flow_chart}. As shown in Fig. \ref{fig:flow_chart}, instead of a feature mapping circuit, we utilise the problem-inspired circuit, which is proposed to generate the dataset with the help of a unitary gate. The proposed method also overcomes the issue of increased dimension/qubits with the qubits in quantum states. The problem-inspired circuit, which generates the quantum states, is further appended with PQC for training and testing. In the following subsection, we briefly discuss the generation of quantum states and the selection of PQC.\par

\subsection{Generation of Dataset}
We use the problem-inspired quantum circuit to generate the dataset. The created circuit is equipped with parametrised two-qubit unitary operators. The entangling power of the applied parametrised two-qubit unitary operators or gates is controlled by varying parameters; these constructed gates can be used to create different classes of entangled multi-qubit quantum states. The use of such parametrised gates provides a unique way to generate datasets. We can construct such operators using the dilation theorem and non-unitary Kraus operators ($K_{i}$) of various noisy quantum channels. The Kraus operators follow relation $\sum_{i}K_{i}^{\dagger}K_{i}=I$. A particular example of the use of Markovian Amplitude Damping (AD) and non-Markovian Random Telegraph Noise (RTN) Kraus operators \cite{Sabale2024_cqc, Sabale2024_Wm} is presented in the article. The dilation process results in a two-qubit unitary operator $U_{i}$, which incorporates noisy channel single-qubit non-unitary Kraus operators. In the case of AD, Kraus operators are 

\begin{equation}\label{k0}
    k_{0} = \begin{pmatrix}
        1 & 0 \\
        0 & \sqrt{n_f}
    \end{pmatrix},
\end{equation}
\begin{equation}\label{k1}
    k_{1} = \begin{pmatrix}
        0 & \sqrt{1-n_f} \\
        0 & 0
    \end{pmatrix},
\end{equation}

where $n_f=1-e^{\alpha \beta}$. In the case of RTN, Kraus operators are expressed as
\begin{align}
    K_{0} &= \sqrt{q_{0}} I , \\
    K_1&= \sqrt{q_{1}} \sigma_{z},
\end{align}
here, $\sigma_{z}$ is a Pauli-Z operator and the coefficients $q_{i}$ associated are of the form 
\begin{equation}
    q_{0}= \frac{1}{2}[1+p(t)] , \quad q_{1}= \frac{1}{2}[1-p(t)].
    \label{eq:prob}
\end{equation}
The $p(t)=\exp(-\gamma t)(\cos{(\omega \gamma t)}+\frac{\sin{(\omega \gamma t)}}{\omega})$ represents a noise function. Following articles \cite{hu2020quantum,PhysRevA.106.022424}, the non-unitary single qubit operator is converted into a two-qubit unitary operator given as
\begin{equation}\label{eq.uni}
    U_i = \begin{pmatrix}
        k_i & D_{i}^{\dagger}\\
        D_{i} & -k_{i}^{\dagger}
    \end{pmatrix},
\end{equation}
and the elements of the resultant matrix are defined as
\begin{equation}
    D_0 = \sqrt{I-k_{0}^{\dagger}k_0},
\end{equation}
\begin{equation}
    D_{0}^{\dagger} = \sqrt{I-k_{0}k_{0}^{\dagger}},
\end{equation}
\begin{equation}
    D_1 = \sqrt{I-k_{1}^{\dagger}k_1},
\end{equation}
\begin{equation}
    D_{1}^{\dagger} = \sqrt{I-k_{1}k_{1}^{\dagger}}.
\end{equation}

The constructed unitary operators ($U_i$), with parameters $\alpha, \beta$ for AD-based Kraus operators, can be further applied to multi-qubit states to generate different classes of quantum states by varying the parameters. This allows us a simple, straightforward way to generate various classes of quantum states. This approach can be easily scaled up for multi-qubit systems. It also allows quantum state data to be fed to a QNN in the form of a single parametrised quantum circuit. This results in a reduction in the effort of encoding quantum states, as well as a reduction in the required resources. Building on this approach, we generate the dataset for three-qubit quantum states using a unitary operator constructed from Kraus operators of RTN and AD noise. The quantum circuits used for generating two-qubit and three-qubit states are shown in Figs. \ref{fig:combine_circuits}A, and \ref{fig:combine_circuits}B. To construct all possible classes of three-qubit quantum states, the unitary operator is applied in various combinations and is depicted in Fig. \ref{fig:combine_circuits}B.

\begin{figure*}[ht!]
    \centering
    \includegraphics[width=\linewidth]{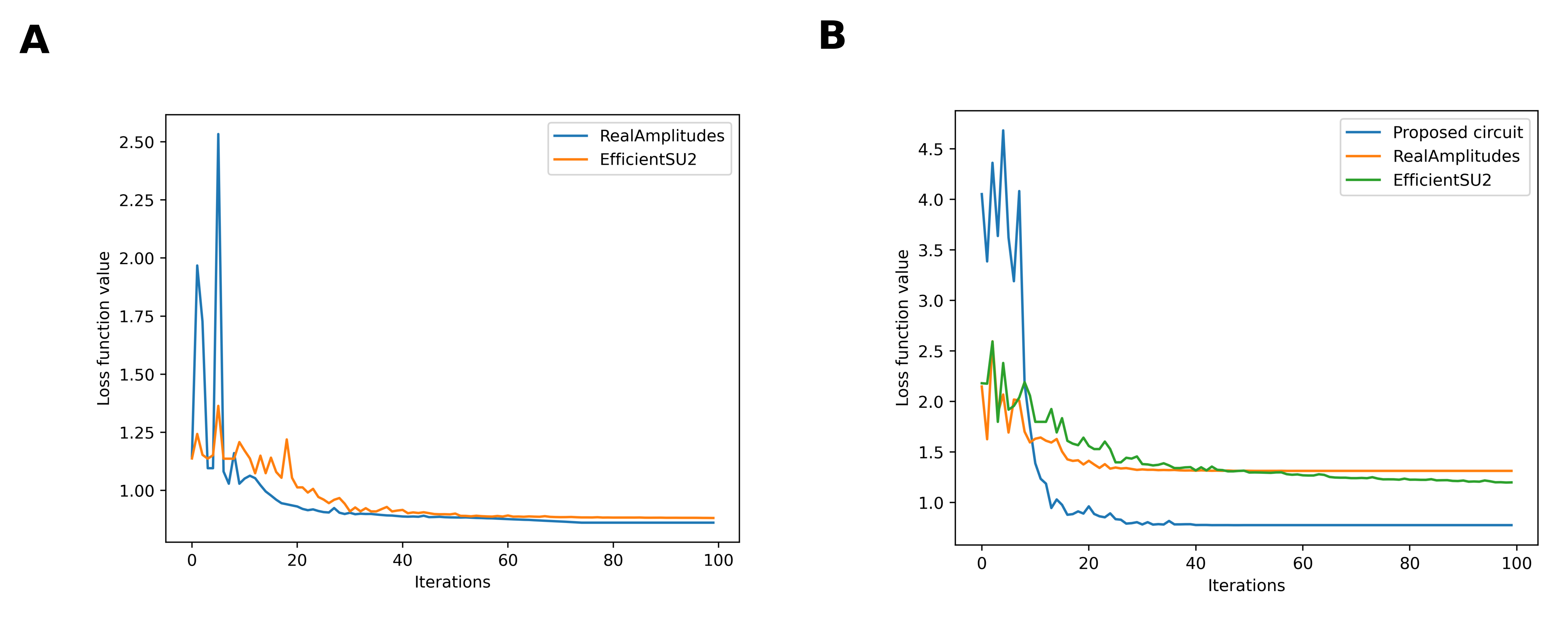}
    \caption{The graphs represent the loss function minimization curve over 100 iterations A) For RealAmplitudes and EfficientSU2 circuits utilized for two-qubit quantum state classification. B) Circuits utilized for three-qubit quantum state classification for the proposed circuit, RealAmplitudes, and EfficientSU2 circuits.}
    \label{fig:loss_functions}
\end{figure*}

\subsection{Training neural networks}
To train the QNN model on the generated quantum states, the states are appended to PQC or Ans\"{a}tz. The PQC circuit is designed based on the ability to get trained while avoiding the condition of barren plateaus. The parameterised gates in the PQC are optimised using classical optimisers, where the initial values of parameters are chosen randomly. Though there are many ways to design an\"{s}atz, this paper has utilised Qiskit's default an\"{s}atz, RealAmplitudes, and EfficientSU2 for two-qubit quantum state classification \cite{qiskit2024}. For three-qubit, in addition to the mentioned an\"{s}atz, we also include a customised an\"{s}atz with Toffoli gates, depicted in Fig. \ref{fig:combine_circuits}C. To achieve the goal of better-optimised parameters for desirable predictions of quantum states, we compute a cost function to minimise the distance between the predictions and actual labels. For this, we use the cross-entropy cost function, 
\begin{equation}
    C({\ket{\psi_{x_{i},\theta}},y_{i}}) = -\sum_{k}y_{k}\log(p_{k}),
\end{equation}
where $k$ is the number of classes in the dataset and $p_{k}$ is obtained by measuring the output state $\ket{\psi_{x_{i},\theta}}$ on the observable $O$. With the computation of the cost function, the task of training the neural network is transferred to minimising the cost function $\theta = argmin_{\theta}C({\ket{\psi_{x_{i},\theta}},y_{i}})$. For the cost function minimisation, we consider a gradient-free optimiser named constrained optimisation by linear approximation (COBYLA) \cite{powell1994direct}. The COBYLA optimiser considers initial parameters to form a simplex to capture the slope of the cost function obtained by
\begin{equation}
    \Delta C = \norm{\theta^{*} - \theta}\leq r,
\end{equation}
where $\theta^{*}$ is any new parameter, replaced with the highest cost value parameter along the slope while staying in the trust region with radius $r$. Therefore, it minimises the cost function $\Delta C$ with the given constraint. However, the random initialisation of parameters can also lead to the condition of a barren plateau \cite{arrasmith2021effect}; therefore, in this study, we will also correlate the impact of barren plateaus on QNN performance.

\begin{figure*}[t!]
    \centering
    \includegraphics[width=\textwidth]{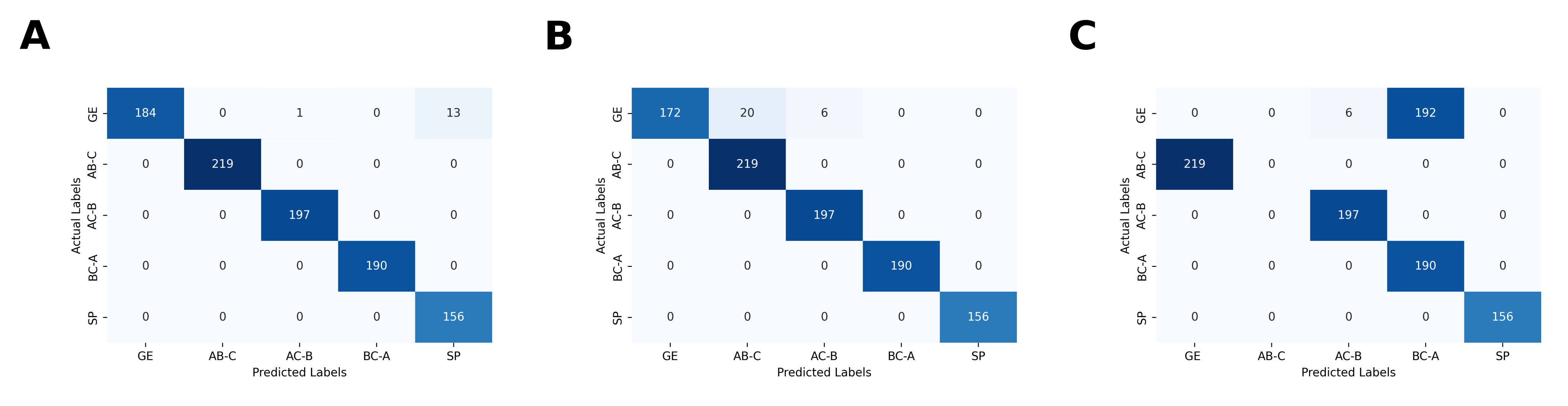}
    \caption{The confusion metrics correspond to the parameterised quantum circuits utilised for three-qubit quantum state classification. The metrics demonstrate the detailed breakdown of predicted classes/labels for the test datasets vs actual labels across different classes. Specifically, A) corresponds to the proposed circuit, B) represents the EfficientSU2 circuit, and C) depicts the RealAmplitudes circuit.}
    \label{fig:confusion_metrics}
\end{figure*}

\section{Numerical Implementation}
To implement a QNN, we first generate a dataset of entangled and separable states for a two-qubit and three-qubit quantum state classification problem. The method enables the generation of an equivalent number of diverse classes for both two-qubit and three-qubit quantum states. Different classes are generated for three-qubit quantum states; here, AB-C, AC-B, and BC-A are bi-separable states, genuinely entangled states (ABC), and separable states (A-B-C). The process of classifying quantum states is further conducted using PQC, also referred to as an\"{s}atz. The selection of an appropriate an\"{s}atz is determined based on factors such as vanishing gradients of cost functions and trainability. For the two-qubit and three-qubit states, we utilise the RealAmplitudes \cite{arthur2022hybrid} and EfficientSU2 \cite{qiskit2024} circuits. To assess the effectiveness of an an\"{s}atz, we compare circuits by analysing the presence of barren plateaus and the number of iterations required for the cost function to converge. To evaluate the circuits for the presence of barren plateaus across the entire parameter space, we uniformly sampled multiple random points within the interval $[-\pi,\pi]$ for both the x-axis and y-axis, creating a grid. The z-axis corresponds to the loss value; here, the loss function is computed by measuring the an\"{s}atz, as given below
\begin{equation}
    L(\theta) = Tr[OU^{\dagger}_{x_{i}}U^{\dagger}_{\theta_{i}}\ket{0}\bra{0}^{\otimes n}U_{\theta_{i}}U_{x_{i}}],
\end{equation}
where $O = I - \ket{0}\bra{0}^{\otimes n}$ and optimised by shifting the parameters throughout the landscape. Fig. \ref{fig:barren_plateaus} illustrates the loss landscape of all an\"{s}atz considered in this study for the classification of two-qubit and three-qubit quantum states. The arrangement of the loss landscapes is as follows: the rows, ordered from top to bottom, represent an\"{s}atz corresponding to two-qubit and three-qubit quantum states, respectively. Generally, barren plateaus are characterised by flat landscapes, while grooves indicate local and global minima. The centre point of each landscape represents the global minimum, corresponding to the optimised parameters for the circuit. As shown in Fig. \ref{fig:barren_plateaus}, for two-qubit circuits, the EfficientSU2 circuit exhibits multiple local minima, whereas the RealAmplitudes circuit demonstrates fewer local minima. However, when these circuits are applied to three-qubit quantum states, the barren plateau phenomenon becomes apparent. Interestingly, modifying the entangling gates within the RealAmplitudes circuit yields significantly improved results, as confirmed by subsequent trainability evaluations. The proposed circuit, incorporating these modifications, is depicted in Fig. \ref{fig:combine_circuits}C. In the following subsection, we will further analyse the models corresponding to accuracy in the classification of the quantum states.

\begin{table}[h!]
    \centering
    \begin{tabular}{c|c|c}
    \hline
         Metrics & RealAmplitudes & EfficientSU2 \\
         \hline
         Accuracy & 99\% &97\%\\ 
         Precision & 99\% &97\%\\ 
         Recall & 99\% &97\%\\ 
         F1-score & 99\% &97\%\\ 
    \end{tabular}
    \caption{A comparison of considered two-qubit quantum circuits on the basis of accuracy, precision, recall, and F1-score.}
    \label{tab:two_qubit_metrics}
\end{table}

\begin{table}[h!]
    \centering
    \begin{tabular}{c|c|c|c}
    \hline
         Metrics & Proposed circuit & RealAmplitudes & EfficientSU2 \\
         \hline
         Accuracy & 99\%& 57\%& 97\%\\
         Precision & 98\%& 49\% & 98\%\\
         Recall & 99\% & 60\%& 97\%\\
         F1-score &98\% &53\%& 97\%
    \end{tabular}
    \caption{A comparison of considered three-qubit quantum circuits on the basis of accuracy, precision, recall, and F1-score.}
    \label{tab:three_qubit_metrics}
\end{table}

\subsection{Quantum state classification}
After selecting the feature mapping circuit and the PQC, we proceeded to train the QNN model using 70\% of the randomly selected generated states for training and the remaining 30\% for testing. Figs. \ref{fig:loss_functions}A illustrates the convergence of the loss function with respect to the number of iterations for two-qubit quantum states. The results show rapid convergence of the loss function near 50 iterations for the RealAmplitudes circuit. However, the convergence remains gradual beyond this point, necessitating additional iterations for full optimisation. In contrast, the EfficientSU2 represents the loss function convergence around 100 iterations. Additionally, we compared the performance of the QNN models based on four evaluation metrics: accuracy, precision, recall, and F1-score. Notably, the RealAmplitudes circuit outperformed the EfficientSU2 circuit across all evaluation metrics, as summarised in Table \ref{tab:two_qubit_metrics}.\par

For the classification of three-qubit quantum states, we evaluated the proposed circuit alongside the RealAmplitudes and EfficientSU2 circuits. As shown in Fig. \ref{fig:loss_functions}B, the loss function for the proposed circuit converges rapidly, stabilising nearly 40 iterations. In comparison, while the loss function for the RealAmplitudes circuit converges more quickly than that of the EfficientSU2 circuit, the final loss value remains high for the RealAmplitudes circuit, and also the corresponding loss landscape, as observed in the barren plateau analysis, is nearly flat. Consequently, as shown in Table \ref{tab:three_qubit_metrics}, the RealAmplitudes circuit achieves an accuracy of 57\%. In contrast, the proposed circuit demonstrates the highest accuracy among all evaluated models. In addition, the proposed circuit achieved the highest accuracy in multi-class classification for distinguishing all possible classes in three-qubit quantum states. A detailed breakdown of the predicted classes for all the PQCs used in this study is presented in Fig. \ref{fig:confusion_metrics}. As illustrated in Fig. \ref{fig:confusion_metrics}A, the proposed circuit successfully predicts every class in the test dataset, with the exception of the generalised entangled state, which is misclassified as a separable class. In contrast, the EfficientSU2 circuit, shown in Fig. \ref{fig:confusion_metrics}B, misclassifies the entangled state as bi-separable states. Similarly, the RealAmplitudes circuit, depicted in Fig. \ref{fig:confusion_metrics}C, fails to classify one of the bi-separable classes correctly and also misclassifies the entangled states. \par
In addition to utilising a unitary operator corresponding to AD noise, we also evaluate the model with RTN-generated datasets. For the two-qubit quantum state classification, the results have shown 100\% accuracy.

\section{Conclusion}
This study addressed the problem of classifying quantum states as entangled or separable using a QNN. In general, quantum state classification is approached using machine learning algorithms and data from randomly generated density matrices of quantum states belonging to different classes. The elements of resulting density matrices are used as features for training algorithms. This approach typically requires a feature space of size $2^n \cross 2^n$-dimension for $n$-qubit quantum state. In this work, we constructed a resource-efficient quantum circuit, named problem-inspired quantum circuit, to generate the two-qubit and three-qubit quantum states with potential scalability to $n$-qubit systems. Notably, the problem-inspired circuit efficiently generated all possible subclasses of three-qubit quantum states. During the classification of quantum states, we observed that the an\"{s}atz, which exhibited a barren plateau, performed less accurately compared to other circuits. Interestingly, for three-qubit quantum states, an increase in the number of Toffoli gates in the an\"{s}atz improved both trainability and classification accuracy. Therefore, these findings conclusively demonstrated the efficiency of the proposed problem-inspired circuit and the customised an\"{s}atz in quantum state classification tasks. This work can be extended to multi-qubit quantum state classification.

% \section*{Code and data availbility}
% The data set involving two-qubit and three-qubit states belonging to different classes is generated using quantum circuits. The used codes and data sets are available in the \href{https://github.com/Thirumalai-97/QuantumStateClassification_UsingQNN.git}{GitHub} repository.

\begin{acknowledgements}
    DS, VBS, and TM express their gratitude to IIT Jodhpur for providing research facilities.
\end{acknowledgements}

\bibliography{apssamp}

\end{document}